\documentclass[12pt]{article}
\pdfoutput=1
\usepackage{graphicx}
\usepackage{xcolor}
\usepackage{amsmath,amssymb,amsfonts}
\usepackage{cite}
\usepackage{multirow}
\usepackage{framed}
\usepackage{slashed}
\usepackage{bbm}
\usepackage{array}
\usepackage{booktabs}
\usepackage{hyperref}
\usepackage{color}
\usepackage{subcaption}

\usepackage[a4paper,margin=2.7cm,bmargin=4.5cm]{geometry}
\renewcommand{\baselinestretch}{1.15}

\makeatletter

\@addtoreset{equation}{section}

\def\asymp#1%
{\mathrel{\raisebox{-.4em}{$\widetilde{\scriptstyle #1}$}}}
\makeatother

\newcommand\fig[1] {Fig.\,{\ref{#1}}}

\newcommand\app[1] {Appendix~\ref{#1}}
\newcommand\tab[1] {Table~\ref{#1}}

\def\beq{\begin{equation}}
\def\eeq{\end{equation}}
\def\bsp#1\esp{\begin{split}#1\end{split}}
\def\bal#1\eal{\begin{align}#1\end{align}}
\def\beeq{\begin{eqnarray}}
\def\eeeq{\end{eqnarray}}

\newcommand\sigmav {\langle\sigma v_\text{M{\o}l}\rangle}

\newcommand\rd   {\ensuremath{\mathrm{d}}}

\newcommand\ri   {\ensuremath{\mathrm{i}}}

\newcommand\rL   {{\mathrm{L}}}

\newcommand\rO   {{\mathrm{O}}}

\newcommand\tZ   {\theta_{Z}}

\newcommand\tW   {\theta_{\mathrm{W}}}
\newcommand\cW   {\cos\tW}

\newcommand\rPl  {\ensuremath{\mathrm{Pl}}}

\newcommand\w[1]{_{\mathrm{#1}}}
\newcommand\TT{\mathrm T}
\newcommand\unit[1]{\,\mathrm{#1}}
\newcommand\GeV{\unit{GeV}}

\newcommand\vev[1]{\langle#1\rangle}

\DeclareMathOperator\gU{\mathrm U}
\DeclareMathOperator\gSU{\mathrm SU}
\newcommand{\UOne}{\gU(1)}
\newcommand\pmat[1]{\begin{pmatrix}#1\end{pmatrix}}
\DeclareMathOperator\diag{\mathrm{diag}}

\def\stars{\strut\leaders\hbox{*}\hfill\strut}
\def\starline{\hfil\strut\hfil\hbox to \textwidth {\stars}\hfil}

\newcommand\MAILTO[1]{\href{mailto:#1}{#1}}

\begin{document}


\begin{titlepage}
\renewcommand{\thefootnote}{\fnsymbol{footnote}}

\mbox{}\par\vskip15mm

\begin{center}
{\Large \bf
Sterile neutrino dark matter in a U(1) extension of the standard model
}
\end{center}
\par \vspace{5mm}

\begin{center}
\textbf{%
  Sho Iwamoto$^1$,
  K\'aroly Seller$^1$\footnote{Corresponding author; email: \MAILTO{karoly.seller@ttk.elte.hu}},
  and
  Zolt\'an Tr\'ocs\'anyi$^{1,2}$
}
\end{center}

\vspace{2mm}

\begingroup
\begin{minipage}[t]{0.9\textwidth}
\centering\renewcommand{\arraystretch}{0.9}
\begin{tabular}{c@{~}l}
$^{1}$ & Institute for Theoretical Physics, ELTE E\"otv\"os Lor\'and University, \\
       & P\'azm\'any P\'eter s\'et\'any 1/A, H-1117 Budapest, Hungary, \\[3mm]
$^{2}$ & MTA-DE Particle Physics Research Group,\\
       & Bem t\'er 18/A, H-4026 Debrecen, Hungary.  
\end{tabular}
\end{minipage}
\endgroup

\par \vspace{15mm}

\begin{minipage}[t]{0.9\textwidth}
\begin{center} {\bf Abstract} \end{center}
\pretolerance 10000
We explore the parameter space of a U(1) extension of the standard model---also called the super-weak model---from the point of view of explaining the observed dark matter energy density in the Universe. The new particle spectrum contains a complex scalar singlet and three right-handed neutrinos, among which the lightest one is the dark matter candidate. We explore both freeze-in and freeze-out mechanisms of dark matter production.
In both cases, we find regions in the plane of the super-weak coupling vs.~the mass of the new gauge boson  that are not excluded by current experimental constraints.
These regions are distinct and the one for freeze-out will be explored in searches for neutral gauge boson in the near future.
\end{minipage}

\end{titlepage}
\clearpage

\setcounter{footnote}{0}
\renewcommand{\thefootnote}{\arabic{footnote}}

\begingroup
\renewcommand{\baselinestretch}{0.5} 
\hrule
\tableofcontents
\vskip .2in
\hrule
\vskip .4in
\endgroup

\section{Introduction}
\label{sec:intro}

The existence of dark matter---one or possibly multiple massive, non-baryonic species of particles---is a well established experimental fact.
The satellite experiments WMAP~\cite{Hinshaw:2012aka} and Planck~\cite{Aghanim:2018eyx} were able to determine the energy density of dark matter through their accurate measurements of the anisotropies in the cosmic microwave background, while numerous other observations were made on the gravitational effects of dark matter, most notably the baryon acoustic oscillations~\cite{Eisenstein:2005su} (see also in the flatness of the galactic rotation curves~\cite{Sofue:2000jx} or through gravitational lensing of galaxies~\cite{Bartelmann:1999yn}, etc.).
However, the particle spectrum in the standard model of particle interactions does not include a viable candidate for dark matter, despite its extensive success. Hence, assuming that a yet unknown particle species is responsible for the effects we observe, an extension of the standard model is required.

Many extensions of the standard model have already been proposed, often focusing on different aspects of new physics. 
In dark matter research the typical approach is to assume a hidden sector of new elementary particles that are sterile under the standard model interactions with a weak connection, called a portal, between the two sectors.
There are three well-known portals in the literature: (i) the vector or gauge boson portal~\cite{Holdom:1985ag}, (ii) the Higgs portal~\cite{Patt:2006fw}, and (iii) the neutrino portal~\cite{Falkowski:2009yz,Macias:2015cna,Blennow:2019fhy}. 
Among the possible extensions of the standard model, an additional U(1) gauge group appears the most economical one, with a new complex scalar (e.g., Refs.~\cite{Sanchez-Vega:2014rka,Rodejohann:2015lca,Bandyopadhyay:2020ufc}) or fermion (e.g., Refs.~\cite{Okada:2010wd,Lindner:2013awa,Biswas:2016bfo,Okada:2020cue}) as dark matter candidate.

A possible fermionic candidate of dark matter is the sterile neutrino \cite{Adhikari:2016bei,Das:2019pua,Seto:2020udg,Belanger:2021slj}. 
It is a neutral lepton beyond the standard model, but unlike known active neutrinos, it does not interact with the standard model particles except through a tiny active-sterile mixing (see, e.g., Ref.~\cite{Boyarsky:2018tvu} for a review).
For example, we may consider a simple scenario, in which the right-handed neutrino dark matter is produced through the active-sterile neutrino mixing (Dodelson-Widrow mechanism)~\cite{Dodelson:1993je}. In this case the same mechanism (i.e., the mixing) is responsible for both the production and the subsequent decay of right-handed neutrinos \cite{Pal:1981rm}, so X-ray observations provide stringent constraints: the sterile neutrino must be lighter than 2\,keV, presuming it explains the total amount of the dark matter \cite{Watson:2011dw,Horiuchi:2013noa,Ng:2019gch}.

Another simple scenario with sterile neutrino dark matter is the aforementioned U(1) extension of the standard model~\cite{Appelquist:2002mw}.
In this case three right-handed neutrinos are necessarily introduced to cancel gauge anomalies.
If we assume the active-sterile mixing is tiny, the lightest right-handed neutrino becomes a sufficiently long-lived dark matter candidate.
As the sterile neutrino couples to U(1), it may be produced by U(1)-mediated processes, such as decays of the extra gauge boson or annihilation of the standard model fermions, which makes the lifetime independent of the production mechanism to avoid various constraints~\cite{Pospelov:2007mp,Pospelov:2008zw}.

In this paper, we consider a U(1)$_z$ extension of the standard model in which the new vector boson $Z'$ and the lightest of the extra right-handed neutrinos $N_1$ have masses ($M_{Z'}$ and $M_1$) much below the electroweak scale \cite{Trocsanyi:2018bkm}. 
The lightest right-handed neutrino satisfies the conditions for being a dark matter candidate, i.e.,~it is sterile under the standard model, and its lifetime can be sufficiently long (its decays are suppressed by the active-sterile neutrino mixing). 
Our aim is to constrain the parameter space of the model by confronting its predictions with the measured dark matter abundance.

The production mechanism of dark matter determines the characteristics of the possible particle physics models. 
Two main scenarios can be distinguished depending on whether the dark matter species reaches equilibrium with the standard model particles or not; the former is called {\em freeze-out} and the latter {\em freeze-in}. 
For a light new mediator, our model allows for correct dark matter abundances assuming both freeze-out and freeze-in mechanisms. 

In the freeze-out mechanism the most studied dark matter candidate particles are the so-called Weakly Interacting Massive Particles (WIMP).
For dark matter particles with masses of order $\rO(100)~$GeV, the popularity of WIMPs is manifest in the fact that they require cross sections similar in magnitude as those in electroweak processes. 
Many extensions of the standard model have natural candidates for such a particle, most notably the lightest supersymmetric particle had been studied extensively~\cite{Jungman:1995df}. 
In spite of the seemingly natural solution, the null results of direct searches for WIMPs have severely constrained the allowed parameter space~\cite{Akerib:2016vxi,Agnese:2017njq,Aprile:2018dbl}. 
On the other hand, we may consider lighter dark matter candidates with freeze-out. 
In this case new constraints arise that limit the mass of the dark matter particles as well as the couplings between the dark sector and the standard model. 
Primary constraints are due to experimental results on Big Bang nucleosynthesis and cosmic microwave background, limiting the mass and annihilation cross sections of new particles \cite{Cyburt:2015mya,Hufnagel:2017dgo,Depta:2019lbe,Boyarsky:2020dzc}. 
From the particle physics side, beam dump and collider searches put limits on the parameter space \cite{Alekhin:2015byh,Kou:2018nap,NA64:2019imj}.
In addition to these, measurements of stellar cooling \cite{Hardy:2016kme} and supernovae \cite{Bollig:2020xdr,Croon:2020lrf} provide constraints for particle physics models with light new mediators in the weak coupling regime.

While freeze-out is experimentally more accessible, the parameter space discussed in the freeze-in scenarios is largely unconstrained due to a very weak coupling between the dark and standard model sectors. 
These dark matter candidates are commonly referred to as Feebly Interacting Massive Particles (FIMP); for a nice review, check Ref.~\cite{Bernal:2017kxu}. 
Freeze-in production of sterile neutrino dark matter have been discussed in e.g., Refs.~\cite{Biswas:2016bfo, Shuve:2014doa,Khalil:2008kp}; see also the white paper~\cite{Adhikari:2016bei}.

In this paper we use natural units throughout, which means that all quantities are measured in powers of GeV.

\section{Particle physics model}
\label{sec:1ParticlePhysicsModel}

We consider an extension of the standard model by a U(1)$_z$
gauge group with particle content and charge assignment shown in Table~\ref{tab:ChargeAssignment} (cf.\ Ref.~\cite{Trocsanyi:2018bkm}).
This {\em super-weak model} is an economical extension of the standard model, designed to explain the origin of (i) neutrino mass and oscillations \cite{Iwamoto:2021wko}, (ii) dark matter (this work), (iii) cosmic inflation and stabilization of the electroweak vacuum \cite{Peli:2019vtp}, (iv) matter-antimatter asymmetry of the universe. 
Our goal in this paper is to constrain the parameter space of the model by assuming that $N_1$ is a candidate for dark matter and its abundance is just sufficient to explain the observed dark matter energy density.
\begin{table}[t]
 \def\sfrac#1#2{#1/#2}
    \centering
    \caption{Particle content and charge assignment of the super-weak model, where $\phi$ and $\chi$ are complex scalars and the others are Weyl fermions. For SU(3)$_\mathrm{c}\otimes$SU(2)$_\mathrm{L}$ the representations, while for U(1)$_y\otimes$U(1)$_z$ the charges ($y$ and $z$) of the respective fields are given. Note that for U(1)$_y$, the eigenvalues of the half hypercharge operator are given.}
    \label{tab:ChargeAssignment}
    \begin{tabular}{ccccc}\toprule
         & SU(3)$_\mathrm{c}$ & SU(2)$_\mathrm{L}$ & U(1)$_y$ & U(1)$_z$ \\
        \midrule
        $Q_\mathrm{L}$ & $\mathbf{3}$ & $\mathbf{2}$ & $1/6$ & $1/6$ \\
        $U_\mathrm{R}$ & $\mathbf{3}$ & $\mathbf{1}$ & $2/3$ & $7/6$ \\
        $D_\mathrm{R}$ & $\mathbf{3}$ & $\mathbf{1}$ & $-1/3$ & $-5/6$ \\
        $L_\mathrm{L}$ & $\mathbf{1}$ & $\mathbf{2}$ & $-1/2$ & $-1/2$ \\
        $N_\mathrm{R}$ & $\mathbf{1}$ & $\mathbf{1}$ & 0 & $1/2$ \\
        $e_\mathrm{R}$ & $\mathbf{1}$ & $\mathbf{1}$ & $-1$ & $-3/2$ \\
        \midrule
        $\phi$ & $\mathbf{1}$ & $\mathbf{2}$ & $1/2$ & 1 \\
        $\chi$ & $\mathbf{1}$ & $\mathbf{1}$ & 0 & $-1$ \\
        \bottomrule
    \end{tabular}
\end{table}

The super-weak model contains three right-handed neutrinos $N_i$, a new scalar $\chi$, and the U(1)$_z$ gauge boson $B'_\mu$ in addition to the particles of the standard model. The $\gSU(2)\w L\otimes\UOne_y\otimes\UOne_z$ symmetry spontaneously breaks into the electromagnetic U(1) by the vacuum expectation values of $\phi_0$ and $\chi$.
As described in Appendix~\ref{sec:appendixA}, the neutral gauge bosons $(W^3_\mu, B_\mu, B'_\mu)$ mix into the massless photon $A_\mu$, the $Z$ boson, and an extra boson $Z'_\mu$, where we consider $M_{Z'}\ll M_Z\simeq 91.2\GeV$.
For the masses of the right-handed neutrinos we assume
$M_1 \ll M_Z \lesssim M_{2,3}$.
Moreover---as the name of the model suggests---the new gauge coupling $g_z$ is taken to be smaller than the weak coupling $g_\rL$, in particular we assume $g_z\ll g_\rL$.

For our purposes, we here present only the couplings between the gauge bosons and leptons; see Ref.~\cite{Trocsanyi:2018bkm} for other aspects of the model.

Considering only the part relevant to the neutral bosons, the covariant derivative is given by
\begin{equation}
    \mathcal{D}_\mu
    \supset -\ri(\mathcal{Q}_{A}A_\mu + \mathcal{Q}_{Z}Z_\mu + \mathcal{Q}_{Z'}Z'_\mu)
    \label{eq:Dmu-neutral}
\end{equation}
with the effective couplings (see Appendix~\ref{sec:appendixA} for derivation)
\begin{subequations}
    \label{eq:effcoups}
\begin{align}
    \label{eq:QA}
    \mathcal{Q}_{A} &= (T_3 + y)|e| = \mathcal{Q}_{A}^\mathrm{SM}\,, \\
    \label{eq:QZ}
    \mathcal{Q}_{Z} &= (T_3\cos^2\tW-y\sin^2\tW)g_{Z^0}\cos\tZ - \zeta g_z\sin\tZ\,,  \\ \label{eq:QZp}
    \mathcal{Q}_{Z'} &= (T_3\cos^2\tW-y\sin^2\tW)g_{Z^0}\sin\tZ + \zeta g_z\cos\tZ\,.
\end{align}
\end{subequations}
The $\gSU(2)_\rL$, $\UOne_y$, and $\UOne_z$ gauge couplings are respectively denoted by $g_\rL$, $g_y$, and $g_z$ and corresponding charges are given by $T_3$, $y$, and $z$.
The Weinberg angle $\tW$ and the $Z$--$Z'$ mixing angle $\tZ$ describe the gauge boson mixing.
The electromagnetic charge is given by $|e|=g_\rL\sin\tW$ as in the standard model, where $g_{Z^0}= g_\rL/\cos\tW$.
A scale dependent effective charge  $\zeta(\mu) = z -\eta(\mu) y$ is also introduced, where $\eta$ is a parameter describing the gauge kinetic mixing.

The smallness of $\tZ$ follows our assumption $g_z\ll g_{Z^0}$ and $M_{Z'}\ll M_Z$:
\begin{equation}
  \tan(2\tZ) = \frac{4\zeta_\phi g_z}{g_{Z^0}} + \rO\left(\frac{g_z^3}{g_{Z^0}^3}\right)\,,
\end{equation}
and the effective couplings are approximately
\begin{align}
 \mathcal{Q}_{Z}
 &= g_{Z^0}\left[T_3\cos^2\tW-y\sin^2\tW + \rO\left(\frac{g_z^2}{g_{Z^0}^2}\right)\right] \, ,
\\
 \mathcal{Q}_{Z'}
 &=g_z\left[q\cos^2\tW(2-\eta) + (z - 2y) + \rO \left(\frac{g_z^2}{g_{Z^0}^2}\right) \right]
\label{eq:QZp-vec}
\end{align}
with $q=T_3+y$ being the electromagnetic charge. 
In particular we see that the weak neutral current reduces to its standard model formula up to small corrections. With the addition that the weak charged current is unaffected by the U(1) gauge extension, a handful of constraints can be trivially evaded by the right choice of parameter space.
Interestingly, one can check that the $Z'$--fermion--fermion interactions are essentially vector-like, with negligible axial vector contribution\footnote{In fact, this property originates in the anomaly-free condition, which requires the $\UOne_z$ charges must be a linear combination of $y$ and $B-L$ with $B$ ($L$) being the baryon (lepton) number.}.
We also emphasize that the $Z'$--neutrino--neutrino interaction is independent of $\eta$, consequentially our results are less affected by $\eta$ (cf.~Appendix \ref{sec:appendixB}).

\section{Dark matter production}

To calculate the evolution of the number density of a dark matter species during the expansion of the early Universe we need to solve the Boltzmann equation (see Ref.~\cite{Kolb:1990vq}). 
In general, we allow for decays and annihilations to affect evolution. 
Let us introduce the dimensionless function for a species $i$ of mass $m_i$
\begin{equation}
    \mathcal{Y}_i(T;m_i,g_i)=\frac{n_i(T;m_i,g_i)}{s(T)}\,,
\end{equation}
called comoving number density, where $n_i$ is its number density and $s(T)$ is the total entropy density. 
The parameter $g_i$ is the internal degree of freedom of species $i$. 
Starting from the Boltzmann equation, we can derive the differential equation for the dark matter candidate $a$,
\begin{subequations}
\label{eq:BoltzmannEquation}
\begin{align}
    \frac{\rd\mathcal{Y}_a}{\rd z} =& \sum_{\{f_1,f_2\}}\sqrt{\frac{\pi}{45}}g^*(\Lambda/z)\frac{m_\rPl\Lambda}{z^2}\langle\sigma_{a+b\to f_1+f_2} v_\text{M{\o}l}\rangle\left[\frac{\mathcal{Y}_a^\text{eq}\mathcal{Y}_b^\text{eq}}{\mathcal{Y}_{f_1}^\text{eq}\mathcal{Y}_{f_2}^\text{eq}}\mathcal{Y}_{f_1}\mathcal{Y}_{f_2} - \mathcal{Y}_a\mathcal{Y}_b\right] \label{eq:BoltzmannAnnihilation} \\
    & +\sum_{\{f'_1,f'_2\}}\sqrt{\frac{45}{4\pi^3}}\frac{g^*(\Lambda/z)}{g_s^*(\Lambda/z)}\frac{m_\rPl z}{\Lambda^2}\langle\Gamma_{a\to f'_1+f'_2}\rangle\left[\frac{\mathcal{Y}_{f'_1}\mathcal{Y}_{f'_2}}{\mathcal{Y}_{f'_1}^\text{eq}\mathcal{Y}_{f'_2}^\text{eq}}\mathcal{Y}_a^\text{eq}-\mathcal{Y}_a \right] \label{eq:BoltzmannDecay}
    \\ \nonumber & + \text{(other processes),}
\end{align}
\end{subequations}
where $b$ may be the same particle as $a$, $m_\rPl=1.2\times 10^{19}\GeV$ is the Planck mass, $v_\text{M{\o}l}$ is the M{\o}ller velocity, and we introduced the dimensionless parameter $z=\Lambda/T$ where $\Lambda$ is an arbitrary mass scale.
The function $g^*(T)$ is the effective relativistic degrees of freedom given by \cite{Gondolo:1990dk}
\begin{equation}
    \label{eq:EffRelDoF}
    g^*(T) = \frac{g_s^*(T)}{\sqrt{g^*_\rho(T)}}\left[1+\frac{T}{3g_s^*(T)}\frac{\rd g^*_s(T)}{\rd T}\right]\, .
\end{equation}
Here $g^*_s(T)$ and $g^*_\rho(T)$ are the relativistic degrees of freedom related to entropy and energy density \cite{Husdal:2016haj}.
For identical initial and final state particles of masses $m_{\rm{in}}$ and $m_{\rm{out}}$, the thermally averaged cross section is \cite{Gondolo:1990dk}
\begin{equation}
    \label{eq:ThermallyAveragedCrossSection}
    \sigmav = \frac{1}{8m_{\rm{in}}^4T [K_2(m_{\rm{in}}/T)]^2}\int_{4\mu^2}^{\infty}\text{d}s~\sigma(s)(s-4m_{\rm{in}}^2)\sqrt{s}K_1\left(\frac{\sqrt{s}}{T}\right)\, ,
\end{equation}
where $\mu=\max(m_{\rm{in}},m_{\rm{out}})$ and $K_i(x)$ is the modified Bessel function of the second kind. 
The thermally averaged decay width for a decaying particle of mass $m$ has a simple analytic form,
\begin{equation}
    \langle\Gamma\rangle = \Gamma\,\frac{K_1(m/T)}{K_2(m/T)}\,.
\end{equation}

There are two dark matter production mechanisms that are potentially interesting in this model: freeze-out and freeze-in. 
In both cases the abundance of dark matter becomes constant below some decoupling temperature $T_\mathrm{dec}$. 
We denote the final abundance as $\mathcal{Y}_\infty$ and connect its value to the dark matter energy density today by
\begin{equation}
    \label{eq:OmegaDMdef}
    \Omega_\mathrm{DM}=6.1\times 10^5\left(\frac{m}{1\,\mathrm{MeV}}\right)\mathcal{Y}_\infty
\end{equation}
for non-relativistic dark matter particles of mass $m$.

From the dark sector particles the only sufficiently stable one is the lightest right-handed neutrino $N_1$, which is our proposed dark matter candidate.
We assume tiny active-sterile neutrino mixing, and focus on the production channels of $N_1$ via annihilations of standard model particles and decays of the massive neutral gauge bosons $Z$ and $Z'$.
The freeze-in and freeze-out production mechanisms of dark matter concern different regions of the parameter space and we consider them separately in the following two subsections.

\subsection{Freeze-out scenario}

In the case of freeze-out dark matter production, the dark sector (at least partially) reaches equilibrium at some temperature $T_0$, which is (much) higher than the mass of the dark matter particle.
The way in which the equilibrium distribution had been achieved is unimportant; the only necessary condition is that it had happened before decoupling.
The freeze-out of a species $i$ means that processes involving $i$ cease to be efficient, compared to the Hubble rate at that time, below some decoupling temperature $T_\text{dec}$, which is usually comparable to $m_i$.
The species leaves equilibrium, and if there are no other processes which would change its number density, it freezes out at a constant value $\mathcal{Y}_\infty$. 
The nature of the mechanism indicates that the relevant processes to consider here are annihilations of dark matter particles to standard model ones.
Decays may have played a role in creating the equilibrium distribution, but for decoupling their role is negligible.

We study the lightest right-handed neutrino $N_1$ with a mass of $M_1=10\text{--}50$\,MeV.
Constraints from Big Bang nucleosynthesis, which requires dark matter to have negligible effects around temperatures $T_\mathrm{BBN}=\rO(0.1)\,$MeV~\cite{Adhikari:2016bei,Depta:2019lbe}, are then avoided.
The $Z'$ mass of $2M_1\leq M_{Z'} < 2m_\mu$ is considered, so $Z'$ decays into electrons, all 3 flavours of active neutrinos, and $N_1$. 
With the choice of these $Z'$ masses it is assured that their abundance has mostly diminished by the onset of nucleosynthesis, and thus their effect will be negligible.
However, for $M_{Z'}>m_\pi$ pion production is kinematically allowed, which would affect the proton-neutron conversion rate \cite{Boyarsky:2020dzc}. In the following we will neglect pion production, as the relevant $Z'$ mass range will turn out to be already excluded by laboratory experiments.

In our set-up, the annihilation of $N_1$ to the standard model particles happens via $s$- and $t$-channel processes with the exchange of a massive gauge boson.
The $t$-channel processes via $W^\pm$ exchange are suppressed by the active-sterile neutrino mixing, which we neglected.
The amplitudes for $s$-channel $Z$- and $Z'$-boson exchange are both proportional to $g_z^2$.
Therefore, as $M_{Z'}\ll M_{Z}$, the annihilation is dominated by $s$-channel $Z'$ exchange.
As shown in the previous section, in the limit of $M_{Z'}\ll M_{Z}$ and $g_z\ll g_{Z^0}$ the coupling between $Z'$ and the fermions is vector-like, and in the massless final state approximation the annihilation cross sections are given by
\begin{align}
    \label{eq:CS1}
    &\sigma(N_1N_1\to e^+ e^-) = \frac{1}{12\pi}\sqrt{1-\frac{4M_1^2}{s}}\,\frac{s}{(s-M_{Z'}^2)^2+M_{Z'}^2\Gamma_{Z'}^2}\,g_z^4\left[\left(\eta-2\right)\cos^2\tW + \frac{1}{2}\right]^2\, , \\
    \label{eq:CS2}
    &\sum_i\sigma(N_1N_1\to\nu_i\nu_i) =     \frac{N_\mathrm{f}}{48\pi}\sqrt{1-\frac{4M_1^2}{s}}\,\frac{s}{(s-M_{Z'}^2)^2+M_{Z'}^2\Gamma_{Z'}^2}\,g_z^4\, ,
\end{align}
where $N_\mathrm{f}=3$ is the number of lepton families in the standard model and $\Gamma_{Z'}$ is the total decay width of $Z'$ given by the sum of
\begin{align}
    \label{eq:DecayRate:Z'ee}
    \Gamma(Z'\to e^+e^-) &= \frac{M_{Z'}}{12\pi}g_z^2\left[\left(\eta-2\right)\cos^2\tW +\frac{1}{2}\right]^2\,, \\
    \label{eq:DecayRate:Z'nn}
    \sum_{i=1}^{N_\mathrm{f}}\Gamma(Z'\to\nu_i\nu_i) &= N_\mathrm{f}\,\frac{M_{Z'}}{96\pi}g_z^2\,, \\
    \label{eq:DecayRate:Z'N1N1}
    \Gamma(Z'\to N_1 N_1) &= \frac{M_{Z'}}{96\pi}g_z^2\left(1-\frac{4M_1^2}{M_{Z'}^2}\right)^\frac{3}{2}\,.
\end{align}

There are three parameters that we have to fix in order to solve the Boltzmann equation: the coupling $g_z$, the neutrino mass $M_1$, and the $Z'$ gauge boson mass $M_{Z'}$. 
We will see later that a more convenient way of parameterization is to use the mass ratio $\xi=M_{Z'}/M_1$ and $M_1$ instead, with the coupling $g_z$.
The dependence on $\eta$ is fairly weak as discussed above.

\begin{figure}[t]
    \centering
    \includegraphics[width=0.8\linewidth]{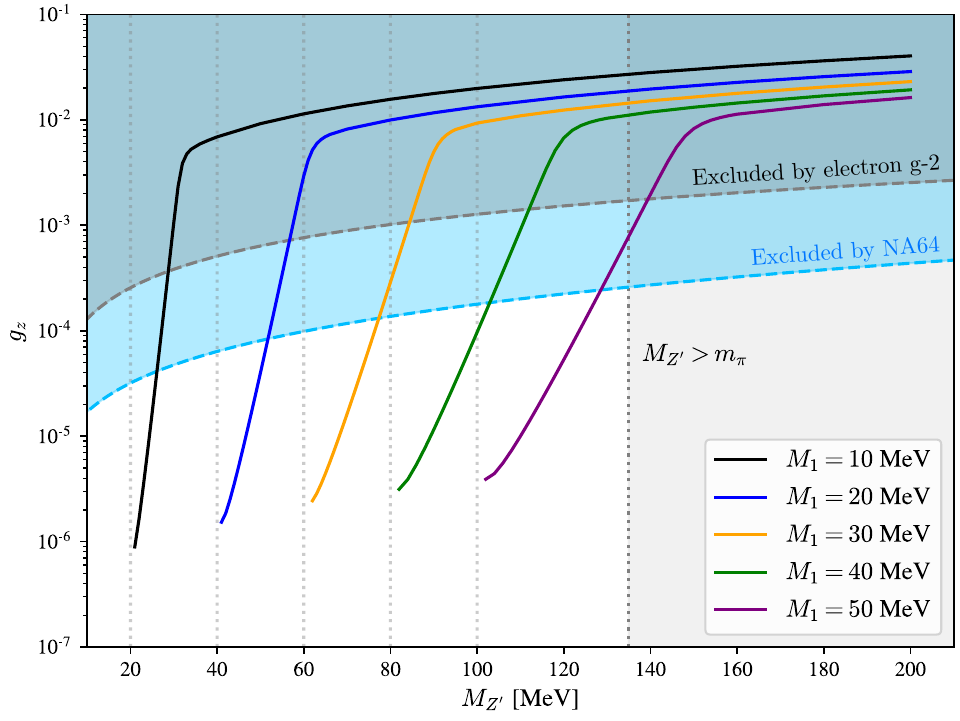}
    \caption{Parameter space for the freeze-out scenario of dark matter production with $\eta=0$. The dark matter particle is assumed to be the lightest right-handed neutrino with mass $M_1$. The required dark couplings $g_z$ reproducing $\Omega_\text{DM}=0.265$ are plotted against the mass of the new gauge boson $Z'$ for various values of the dark matter mass. The shaded region of the parameter space is excluded by the $a_e$ bound (dashed gray) obtained from the U(1)$_z$ contribution to the electron anomalous magnetic moment \cite{Morel:2020dww}, and the NA64 bound (dashed light-blue) obtained from missing energy searches \cite{NA64:2019imj}. The lightly shaded region $M_{Z'}>m_\pi$ is not excluded, but it may be in conflict with the observed proton-to-neutron ratio \cite{Boyarsky:2020dzc}, which we do not consider in detail in this paper since the relevant couplings for $M_{Z'}\gtrsim 130$~MeV are already ruled out by NA64.}
    \label{fig:ResonantFreezeOutParamSpace}
\end{figure}

In the freeze-out mechanism, the relic density of dark matter is inversely proportional to the annihilation cross section of dark matter, $\sigmav$. 
This puts limitations on the parameter space due to overproduction of dark matter for small $\sigmav$. 
From Eq.~\eqref{eq:OmegaDMdef} we find
\begin{equation}
    \label{eq:OmegaWIMPrelation}
    \Omega_\mathrm{DM}\propto \frac{M_1}{g_z^4\langle\tilde{\sigma}v_\text{M{\o}l}\rangle_\mathrm{dec}}\,,
\end{equation}
where $\langle\tilde{\sigma}v_\text{M{\o}l}\rangle_\mathrm{dec}$ is $\sigmav/g_z^4$ at $T=T_\mathrm{dec}$. 
For example, for $M_1=10\,$MeV and $\xi=5$, a coupling of order $g_z=\rO(10^{-2})$ is required for $N_1$ to reproduce the measured dark matter density, $\Omega_\mathrm{DM}=0.265$. However, this value of a new gauge coupling is already ruled out by experiments (see Fig.~\ref{fig:ResonantFreezeOutParamSpace}).
There are two ways in which $g_z$ can be decreased, while keeping $\Omega_\mathrm{DM}$ fixed: either by decreasing $M_1$, or by increasing $\langle\tilde{\sigma}v_\text{M{\o}l}\rangle_\mathrm{dec}$. 
The former option is not favored by Big Bang nucleosynthesis, thus we need to use the latter by exploiting the resonance amplification of the cross section at $M_{Z'}\approx 2M_1$. 

We focus on the parameter space where the resonance gives contribution to the thermally averaged cross section, which corresponds to $M_{Z'}\lesssim 4 M_{1}$ (see \app{sec:appendixB}).
Using a dimensionless integration variable $\lambda=\sqrt{s}/M_{Z'}$ (i.e., the resonance is at $\lambda=1$) we define the bounded thermally averaged cross section as
\begin{equation}
    \label{eq:sigmavEta}
    \sigmav'(u) = \frac{Q(\eta)g_z^4}{4M_{Z'}^2K_2(z)^2}\xi^3 z\int_{2/\xi}^u\rd\lambda\,\frac{\lambda^6}{(\lambda^2-1)^2+\gamma^2}\left(1-\frac{4}{\xi^2\lambda^2}\right)^\frac{3}{2}K_1(\xi z \lambda)\,,
\end{equation}
where $\gamma=\Gamma_{Z'}/M_{Z'}$ and 
\begin{equation}
    \label{eq:Qdef}
    Q(\eta) = \frac{1}{48\pi}\left[4\left(\left(\eta-2\right)\cos^2\tW + \frac{1}{2}\right)^2+3\right].
\end{equation}
The thermally averaged cross section defined in Eq.~\eqref{eq:ThermallyAveragedCrossSection} is $\sigmav=\sigmav'(\infty)$. 
The integral \eqref{eq:sigmavEta} is numerically well defined for large values of $\xi$ because the Bessel function suppresses the resonance\footnote{For large values of $x$, the Bessel function is exponentially small, $K_1(x)\sim\exp(-x)/\sqrt{x}$.}. 
However, now we will consider $\xi\gtrsim 2$, i.e.,~include the contribution of the $Z'$ resonance.
For small values of $\gamma$, we can use the representation of the Dirac-delta distribution,
\begin{equation}
    \label{eq:DiracRep}
    \lim_{\gamma\to 0}\frac{\gamma}{x^2+\gamma^2}=\pi\delta(x)\, ,
\end{equation}
which is an adequate approximation in the integral as long as the major contribution is due to this resonance.
We break up the integral into two terms: the resonant contribution (represented by the Dirac-delta) and the low temperature or high-$\xi$ contribution (i.e., the resonance is suppressed by $K_1(\xi z)$). 
By substitution of Eq.~\eqref{eq:DiracRep} into Eq.~\eqref{eq:sigmavEta} at $u\to\infty$, we find the resonant contribution to be the analytic expression
\begin{equation}
    \label{eq:sigma_resonant}
    \sigmav_\text{res}= \frac{Q(\eta)g_z^4}{4M_1^2K_2(z)^2} \frac{\xi^3 z\pi}{2\gamma}\left(1-\frac{4}{\xi^2}\right)^\frac{3}{2}K_1(\xi z)\,.
\end{equation}

\begin{figure}[t]
    \centering
    \includegraphics[width=0.7\linewidth]{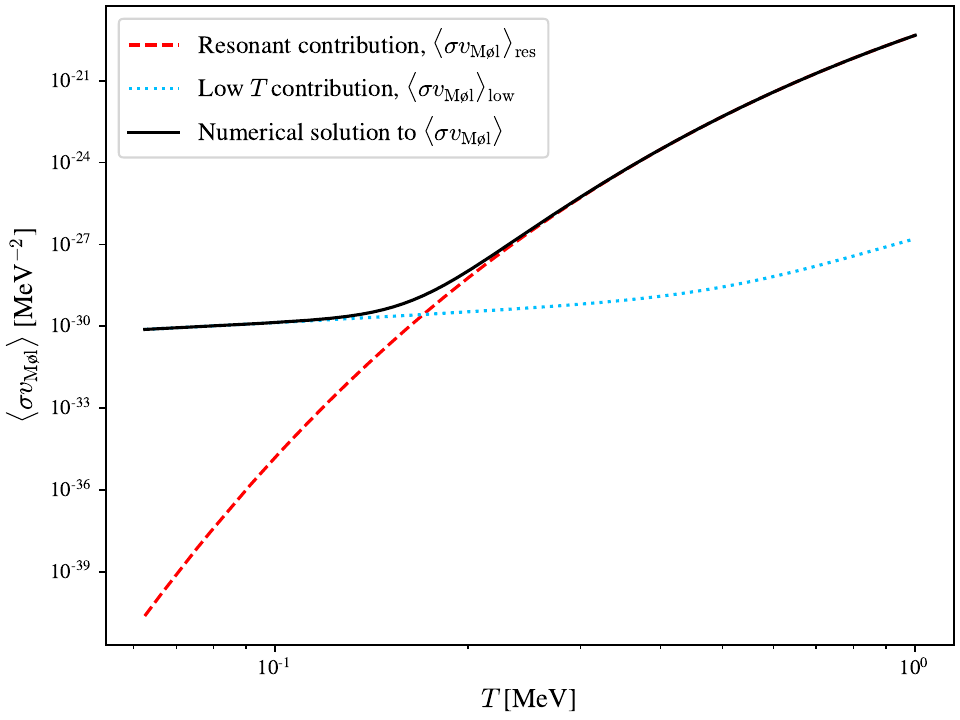}
    \caption{The thermally averaged cross section (solid black) obtained from careful integration of Eq.~\eqref{eq:sigmavEta} numerically. The separate contributions of the resonant (dashed red, cf.~Eq.~\eqref{eq:sigma_resonant}) and the low temperature (dotted blue, cf.~Eq.~\eqref{eq:sigmalow}) part are shown to reproduce the numerical result at their respective regions of validity. In this figure we used the parameters $M_1=10\,$MeV, $M_{Z'}=30\,$MeV ($\xi=3$), $g_z=10^{-2}$, and $\eta=0$.}
    \label{fig:sigmav}
\end{figure}

For the low temperature part the usual approximation of neglecting $\lambda^2$ in the denominator when compared to 1 (i.e.,~the assumption of $s\ll M_{Z'}^2$) is incorrect because $\lambda\geq 2/\xi\sim 1$. What we can do rather is introducing a cutoff to the integral just below the resonance, 
\begin{equation}
    \label{eq:sigmalow}
    \sigmav_\text{low} = \sigmav'(1-\varepsilon)\,.
\end{equation}
The value of $\varepsilon$ does not really matter as long as $\varepsilon\ll 1$, but to follow the reasoning given above it should be chosen as $\varepsilon\sim\gamma$.
This will naturally undershoot the full integral for large temperatures (or for small $\xi$ values), but will converge nicely in the low temperature limit, and is computationally simple.
The total thermally averaged cross section at arbitrary temperature is then approximated as the sum of the two contributions,
\begin{equation}
    \label{eq:sigma_parts}
    \sigmav \simeq \sigmav_\text{low} + \sigmav_\text{res}\,.
\end{equation}
An example of this separation of the thermally averaged cross section is given in Fig.~\ref{fig:sigmav}. We have checked that the integral in Eq.~\eqref{eq:sigmavEta} when computed numerically (shown by the solid black line) is indeed reproduced by the approximation Eq.~\eqref{eq:sigma_parts} as long as the numerical integration was sufficiently stable.

\begin{figure}[t]
    \centering
    \includegraphics[width=0.7\linewidth]{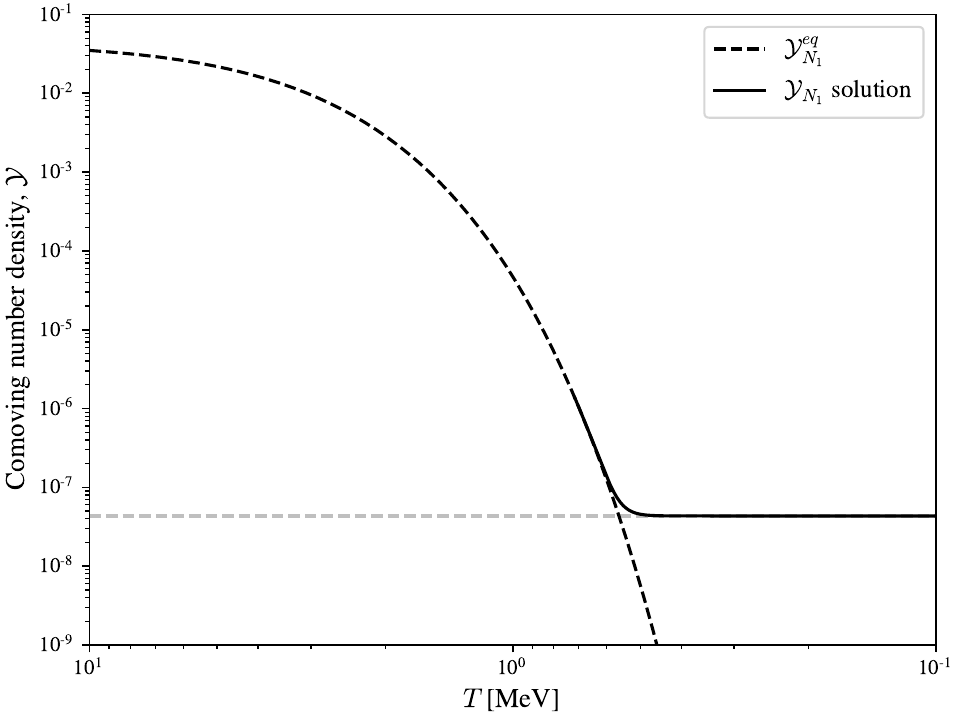}
    \caption{Example solution (solid black) to the Boltzmann equation in the freeze-out case. At high temperatures the solution follows the equilibrium comoving number density (dashed black), while at low temperatures the dark matter decouples, and a non-zero relic density is frozen out. The horizontal line indicates the relic density corresponding to $\Omega_\text{DM}=0.265$, $M_{Z'}=30\,$MeV, $M_1=10\,$MeV, $g_z= 1.06\cdot 10^{-3}$, $\eta=0\,$.}
    \label{fig:FreezeOutExample}
\end{figure}

Substituting Eq.~\eqref{eq:sigma_parts} into Eq.~\eqref{eq:BoltzmannAnnihilation} and using $\Lambda=M_1$ as the relevant mass scale of the problem, we solved the differential equation numerically down to low temperatures around $T_1\simeq M_1/100$ where the solution can be considered a constant, $\mathcal{Y}_\infty$.
The initial condition is simply given by the equilibrium comoving number density for the lightest right-handed neutrino, while the starting temperature can be chosen to be around $T_0\simeq M_1/10$. One such solution is presented in Fig.~\ref{fig:FreezeOutExample}.

In Fig.~\ref{fig:ResonantFreezeOutParamSpace} we see the parameter space that reproduces $\Omega_\mathrm{DM}=0.265$.
For each dark matter mass $M_1$, the resulting curve is clearly divisible into two parts. 
Due to the resonant annihilation of neutrinos at $\xi\simeq 2$, we see a steep drop in the required couplings at lower $Z'$ masses. 
Two constraints are shown as shaded excluded regions. 
The gray comes from the constraints obtained on the anomalous magnetic moment of the electron ($\delta a_e = a_e(\mathrm{exp.})-a_e(\mathrm{theory})<-3.4\cdot 10^{-13}$ at 2$\sigma$ confidence) \cite{Morel:2020dww}. 
The light blue, more restrictive constraint, is from the NA64 experiment \cite{NA64:2019imj} which looks for missing energy in bremsstrahlung processes due to dark photon creation. 
We translated these constraints to the parameters of our model (cf. Ref.~\cite{Ilten:2018crw}); 
the details are described in Appendix~\ref{appendix:C}. In addition, we have indicated the region where pion production is kinematically allowed via $Z'$ decays. 
While it is not strictly excluded, in this region the existence of $Z'$ may modify the proton-to-neutron ratio at the onset of nucleosynthesis due to the appearance of pion-enhanced proton-neutron conversion \cite{Boyarsky:2020dzc}.
Since the relevant coupling $g_z$ required for $N_1$ to reproduce the dark matter density is already excluded for these $Z'$ masses, we do not go into detail with dealing with these constraints.
We also mention that supernova luminosity arguments constrain the model for light $Z'$ bosons \cite{Croon:2020lrf}, however the excluded region lies well below the parameter space shown here\footnote{Detailed analysis of supernova constraints is complicated due to $Z'$ bosons being coupled to the electrons which are highly degenerate within the supernova. Approximate calculations following Ref.~\cite{Croon:2020lrf} using the simulation results of Ref.~\cite{Bollig:2020xdr} indicate that the supernova measurements do not constrain our model in the region of the parameter space discussed here.}. We further mention that lifetime constraints on $Z'$ could be relevant even below the pion mass \cite{Boyarsky:2020dzc}. However, we have checked that the lower bound on $g_z(M_{Z'})$ provided by this argument is below those of the supernova constraints, as such they are not relevant in our model.

We find that for the super-weak model, the coupling range $g_z \in [10^{-6},10^{-4}]$ is not excluded for the resonant production of sterile neutrino dark matter via the freeze-out mechanism.
The parameter region will be searched for in near future by experiments such as Belle II~\cite{Kou:2018nap}, LDMX~\cite{Akesson:2018vlm}, and NA64~\cite{Gninenko:2300189} (cf.\ Ref.~\cite{Battaglieri:2017aum}).

\subsection{Freeze-in scenario}

Contrary to freeze-out, in the freeze-in case of dark matter production, the dark matter candidate species is never in chemical equilibrium with the rest of the cosmic plasma. This is only possible if we assume that the initial abundances of the dark sector particles can be taken zero\footnote{In reality we do not have to be this strict with the choice of initial condition. The relic density $\mathcal{Y}_\infty$ for dark matter is independent of the initial choice for the densities at $T_0$ as long as $\mathcal{Y}_\text{DM}(T_0)/\mathcal{Y}_\infty\ll 1$.} at some early time, i.e.,~after inflation. While such an assumption may appear ad hoc, we cannot exclude it a priori, so we follow up on this possibility. Given that the interactions between the standard model and the dark sector are heavily suppressed by a small coupling, it is possible that the dark matter species never reaches chemical equilibrium before its interactions have ceased.

In the freeze-in mechanism dark matter is produced mainly via decays of heavier particles, while production via annihilations can usually be neglected due to the requirement of very tiny---often called feeble---couplings ($g_z\lesssim 10^{-10}$). 
The production is then qualitatively very simple, the decaying particles will vanish completely, a fraction of them (as given by the relevant branching ratio) producing dark matter particles. 
After the decaying heavy particles have vanished, there are no other processes which could change the comoving number density of the dark matter candidates (provided that they are sufficiently stable), and we obtain a value of the relic density $\mathcal{Y}_\infty$ from which we can calculate $\Omega_\text{DM}$.

The natural candidate for dark matter in our model is the lightest right-handed neutrino. 
Assuming tiny active-sterile mixing, the only vertices of the right-handed neutrino are with $Z$, the two scalars of the theory, and the new gauge boson $Z'$. 
The $Z$--$N_1$--$N_1$ vertex is suppressed by the smallness of the $Z-Z'$ mixing angle $\tZ$, so the decay rate $\Gamma(Z\to N_1N_1)\propto g_z^4$ can be neglected.
Similarly, for an $M_1=\rO(10)$~keV scale neutrino the scalar coupling is suppressed by the ratio $M_1/M_{Z'}\ll 1$ along with the coupling $g_z$, and thus it is also negligible in our discussion (also the scalars vanish at high temperatures, and are unable to produce meaningful abundances of sterile neutrinos even if their decay rates were quantitatively relevant). 
The decay rate of $Z'$ into right-handed neutrinos was given in Eq.~\eqref{eq:DecayRate:Z'N1N1}. 
Since $\Gamma(Z'\to N_1N_1)\propto g_z^2$ this channel is the least suppressed from the three, and the relative smallness of the $Z'$ mass allows a long time ($T_\mathrm{dec}\sim M_{Z'}$) for neutrinos to be created via these decays. 

We may subject all dark sector particles---in particular $Z'$---to the same condition of negligible initial abundance, as the right-handed neutrinos. 
It follows that we have to solve a coupled system of two Boltzmann equations for the out-of-equilibrium densities of $Z'$ and $N_1$ as well. 
However the former can be solved without the inclusion of the latter; if the densities of the final state particles are much smaller than the equilibrium values (i.e.,~$\mathcal{Y}_{N_1}\ll \mathcal{Y}_{N_1}^\text{eq}$), the reverse process can be ignored, and $\mathcal{Y}_{N_1}$ does not appear in the Boltzmann equation for $Z'$ (cf.~Eq.~\eqref{eq:BoltzmannDecay}). In our case this means that while the $Z'$ bosons will decay into right-handed neutrinos, the reverse process is extremely unlikely and it is neglected.

In the freeze-in case we will continue to consider the $Z'$ gauge boson mass range $20\,\mathrm{MeV}<M_{Z'}<200\,\mathrm{MeV}$, however in this case $N_1$ will be taken lighter, around $M_1=\mathrm{O}(10)\,$keV. 
The Big Bang nucleosynthesis constraints are evaded due to $N_1$ having abundances much smaller than the equilibrium and their interactions with standard model particles are feeble. 
The relevant decay rates are the same as those listed in Eqs.~\eqref{eq:DecayRate:Z'ee}--\eqref{eq:DecayRate:Z'N1N1}.

We write the Boltzmann equation Eq.~\eqref{eq:BoltzmannDecay} with $\Lambda=M_{Z'}$ as the relevant scale because freeze-in concludes after the $Z'$ bosons have depleted around $T\simeq 0.1 M_{Z'}$:
\beq
\bsp
    \frac{\rd\mathcal{Y}_{Z'}}{\rd z} &= -\sqrt{\frac{45}{4\pi^3}}\frac{m_\mathrm{Pl}z}{M_{Z'}^2} \frac{g^*(M_{Z'}/T)}{g_s^*(M_{Z'}/T)} \frac{K_1(z)}{K_2(z)} 
    \\ &\times
    \bigg[ \Big(\Gamma(Z'\to e^+e^-) + \sum_{i=1}^3\Gamma(Z'\to\nu_i\nu_i) \Big) \Big(\mathcal{Y}_{Z'} - \mathcal{Y}^\mathrm{eq.}_{Z'}(z)\Big) 
    + \Gamma(Z'\to N_1N_1)\mathcal{Y}_{Z'}
    \bigg],
\\
    \frac{\rd\mathcal{Y}_{N_1}}{\rd z} &= \sqrt{\frac{45}{4\pi^3}}\frac{m_\mathrm{Pl}z}{M_{Z'}^2}\frac{g^*(M_{Z'}/T)}{g_s^*(M_{Z'}/T)}  \frac{K_1(z)}{K_2(z)}\Gamma(Z'\to N_1N_1)\mathcal{Y}_{Z'}\,.
\label{eq:Boltzmannfreezein}
\esp
\eeq
The freeze-in scenario has significantly larger parameter space than we have in freeze-out due to the dark matter species not being in equilibrium at early times. 
There are five parameters: (i) $M_1$ the mass of the lightest neutrino, (ii) $M_{Z'}$ the mass of the new gauge boson, (iii) $g_z$ the new gauge coupling, (iv) $T_\mathrm{rh}$ the reheating temperature, and (v) $\mathcal{Y}_{N_1}(T_\mathrm{rh})$ the initial abundance of neutrinos. 
We mentioned that the differential equation is rather rigid against changing the initial abundance as long as it is kept relatively low compared to the relic abundance. 
A similar statement holds for the initial temperature as well, the relic density is unchanged as long as $T_\mathrm{rh} \gg M_{Z'}$. 
We fix $T_\mathrm{rh}$ and $\mathcal{Y}_{N_1}(T_\mathrm{rh})$ at arbitrary values satisfying these conditions, noting that the results will not depend on them. 
The mass of the lightest neutrino is a trivial parameter, i.e.,~it does not affect the solution of the differential equation, and it only appears in the relation between the relic density and the dark matter density parameter, Eq.~\eqref{eq:OmegaDMdef}. 
As $\Omega_\text{DM}\propto M_1\mathcal{Y}_\infty$, for a minimal value of $M_1$ we can set an upper bound on the coupling reproducing dark matter densities for a given $M_{Z'}$, since $\mathcal{Y}_\infty\propto g_z^2$.

\begin{figure}[t]
    \centering
    \includegraphics[width=0.67\linewidth]{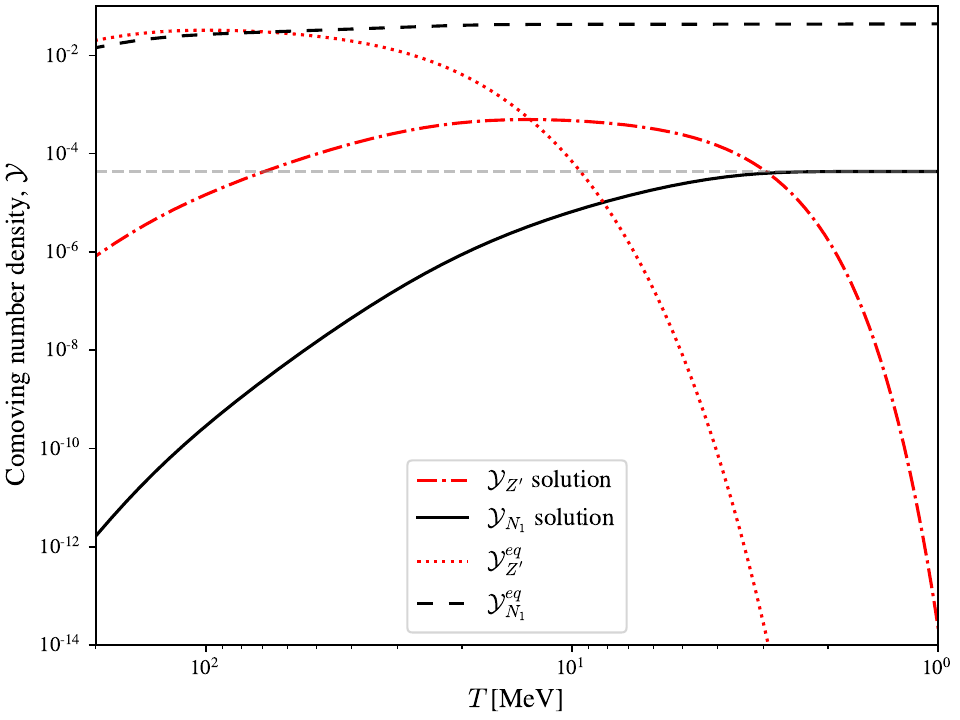}
    \caption{Example solution to the coupled Boltzmann equations in the freeze-in scenario.  For this figure, $M_{Z'} = 100\,$MeV, $g_z\simeq 7.1\cdot 10^{-11}$ and $\eta=0$ were used. The plotted horizontal dashed line corresponds to dark matter densities of $N_1$ particles with mass $M_1=10\,$keV.}
    \label{fig:FreezeInExample}
\end{figure}

We solve the coupled differential equations in Eq.~\eqref{eq:Boltzmannfreezein} for different values of the parameters $M_1$, $M_{Z'}$, and $g_z$. 
An example solution for $M_1=10\,$keV is shown in \fig{fig:FreezeInExample}, where $\mathcal{Y}_{N_1}$ (solid black line) reaches the desired dark matter abundance (dashed gray line).
The comoving number density of $Z'$ bosons (dot-dashed red) increases until it intersects with the equilibrium curve (dotted red), afterwards their decays to neutrinos and electrons dominate. 
Right-handed neutrinos are produced until all of the $Z'$ bosons have depleted, then their abundance becomes constant. 
The equilibrium density for neutrinos (black, dashed) is much larger than the solution throughout the evolution.

\begin{figure}[t]
    \centering
    \includegraphics[width=0.67\linewidth]{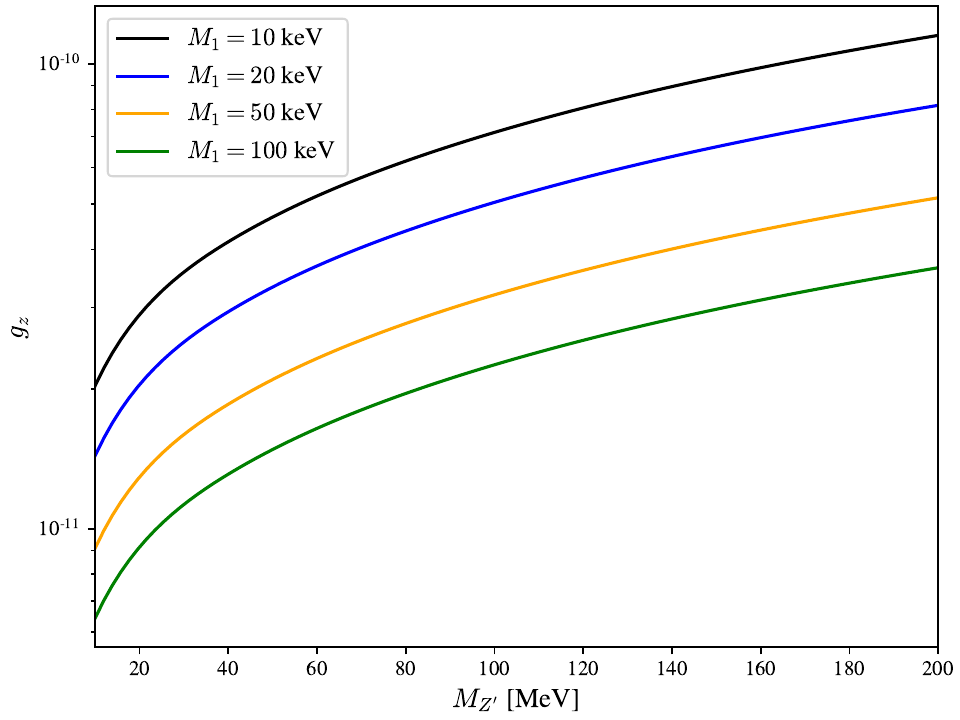}
    \caption{Parameter space for the freeze-in scenario with $\eta=0$. The couplings reproducing dark matter density $\Omega_\text{DM}=0.265$ are shown as a function of the $Z'$ mass for several different values of the dark matter mass $M_1$.}
    \label{fig:FreezeInParameterSpace}
\end{figure}

We explore the parameter space in Fig.~\ref{fig:FreezeInParameterSpace} for $g_z$ and $M_{Z'}$ at several values of $M_1$, while fixing $\eta=0$. 
The couplings reproducing dark matter density $\Omega_\text{DM}=0.265$ are shown for different values for the dark matter mass $M_1$. 
Larger values $M_1$ require smaller couplings because $\Omega_\text{DM}\propto M_1g_z^2$, with typical range in $[10^{-11},10^{-10}]$.
On the other hand, for increasing $M_{Z'}$ the required couplings are also increasing.
Even though the decay rates are proportional to $M_{Z'}g_z^2$, the maximum abundance of $Z'$ bosons is smaller for larger $M_{Z'}$ because there is less time for them to be created ($Z'$ bosons will completely decay by $T\sim 0.01 M_{Z'}$). 

For such a small coupling, direct measurement of this interaction is not possible at present or in the near future. However, there are a number of ways in which the extra particles can be indirectly seen. Measurements regarding Big Bang nucleosynthesis (or the cosmic microwave background) constrain the number of effective relativistic fermionic degrees of freedom $N_\mathrm{eff}$ to lie around its standard model value \cite{Hernandez:2014fha,Vincent:2014rja}. The introduction of new particles which are relativistic and abundant around the times of nucleosynthesis would increase the value of this number, leading to modified cosmological history which does not agree with current measurements. 
We find that in the freeze-in scenario, the sterile neutrino $N_1$ contributes $\Delta N_\mathrm{eff}=\rO(10^{-1}\text{--}10^{-2})$, which is of the order of current experimental uncertainties. 
Future missions are aimed at refining the CMB measurement, thus providing new constraints and an indirect test to our model \cite{Abazajian:2019eic}.
Measurements of stellar cooling also provide stringent constraints, but only for very light mediators, which is below what we consider here \cite{Hardy:2016kme}.

Signals of this scenario may be provided by observations of future supernova explosions.
As studied in Refs.~\cite{Chang:2016ntp,DeRocco:2019njg} for the dark photon models, $Z'$ may induce excessive cooling in the explosion or excessive gamma-ray emission.
With a crude estimation\footnote{
Unlike dark photons, the $Z'$ boson couples to neutrinos and neutrons, for which the channels $\nu\bar\nu\to Z'$ (neutrino coalescence) and $pn\to pnZ'$ (bremsstrahlung from neutrons) contribute to the cooling as well. These processes also result in increased gamma-ray emission, which are however reduced due to the decay channels into neutrinos.
Ignoring these extra effects, we have estimated the constraints from SN1987A and found that, for $10\,\mathrm{keV}<M_1<100\,\mathrm{keV}$, our freeze-in model with $M_{Z'}>40\,$MeV is still allowed.
},
we expect that a small part of the parameter space shown in Fig.~\ref{fig:FreezeInParameterSpace} is constrained by the observation of SN1987A: cooling constraints may provide some exclusion in the upper-left region of the figure, while excessive photon production may do so in the lower-left.
Further analyses are however left for future dedicated studies.

\section{Conclusions}

In this work we have investigated the possible sterile neutrino dark matter production in the framework of a U(1)$_z$ extension of the standard model, also called the super-weak model. 
We have shown that both freeze-in and freeze-out production mechanisms are equally viable to reproduce the observed dark matter energy density although with very different regions in the parameter space regarding the mass of the dark matter particle and the value of the super-weak coupling.

Considering light $Z'$ bosons in the mass range of $M_{Z'}=20$--$200$\,MeV, we have concluded that freeze-out of right-handed neutrinos generally results in over abundance, which can be avoided by considering {\em resonant production} when $M_{Z'}\approx 2M_1$. Such production of $\rO(10)$\,MeV scale right-handed neutrinos is efficient already at small couplings, evading strong constraints due to the experimental upper bounds provided by the measurement of the anomalous magnetic moment of the electron, as well as searches for invisible decays of dark photons by the NA64 experiment.  The lower limit on the mass of $N_1$---and by virtue of the resonant production condition on $M_{Z'}$---is set by Big Bang nucleosynthesis.

In the freeze-in case $M_1=\rO(10)$\,keV scale sterile neutrinos were considered with similarly light $Z'$ bosons as in freeze-out.
We found that they can be responsible for the observed dark matter energy density provided that their coupling is feeble, falls into the range $g_z=10^{-11}$--$10^{-10}$. The parameter space of the freeze-in scenario is constrained by astrophysical observations, mainly due to cooling effects and gamma-ray production in supernova explosions.

The parameter region motivated in the freeze-out case will be explored by Belle II~\cite{Kou:2018nap}, LDMX~\cite{Akesson:2018vlm}, and NA64~\cite{Gninenko:2300189} experiments (cf.\ Ref.~\cite{Battaglieri:2017aum}).
Meanwhile, the freeze-in scenario suggests gauge coupling $g_z$ far smaller than the reach of such particle physics experiments; cosmological and astrophysical constraints---such as an observation of a new supernova---will be required to test the scenario.

\subsubsection*{Acknowledgments}

We are grateful to members of the ELTE PPPhenogroup (pppheno.elte.hu/people) for useful discussions.
This work was supported by grant K 125105 of the National Research,
Development and Innovation Fund in Hungary.  

\appendix

\section{Gauge sector of the super-weak model}
\label{sec:appendixA}

In this appendix, we review the gauge sector of the $\UOne_z$ model.
We first discuss the mixing of $\UOne_y$ and $\UOne_z$ in detail, following Ref.~\cite{delAguila:1988jz}, and derive the effective couplings \eqref{eq:effcoups} and the gauge boson masses.

We start from a general form of the gauge sector in which the gauge kinetic terms and the covariant derivatives relevant for the two $\UOne$ symmetries are given by
\beq
\bsp
\mathcal L&\supset -\frac14 F^{\mu\nu}F_{\mu\nu} - \frac14F'^{\mu\nu}F'_{\mu\nu},\\
 \mathcal{D}^{\mathrm{U}(1)}_\mu &= -\ri \begin{pmatrix}
    y & z
    \end{pmatrix}
    \begin{pmatrix}
    g_{yy} & g_{yz} \\
    g_{zy} & g_{zz}
    \end{pmatrix}
    \begin{pmatrix}
    B_\mu \\ B'_\mu
    \end{pmatrix}\,,
\esp
\eeq
where $y$ and $z$ are the $\UOne_y$ and $\UOne_z$ charges of the corresponding particle and $B_\mu$ and $B'_\mu$ are the $\UOne_y$ and $\UOne_z$ gauge bosons, respectively. We have chosen the basis\footnote{
This basis is used in, e.g., \texttt{SARAH}~\cite{Staub:2008uz}.}
in which the gauge-field strength, $F_{\mu\nu}$ and $F'_{\mu\nu}$, do not mix, while the couplings are given by a $2\times 2$ coupling matrix, which we parameterize as
\begin{equation}
    \mathbf{g}\equiv\begin{pmatrix}
        g_{yy} & g_{yz} \\
        g_{zy} & g_{zz}
    \end{pmatrix} = 
    \begin{pmatrix}
        g_y & -\eta g_z \\
        0 & g_z
    \end{pmatrix}
    \begin{pmatrix}
        \cos\epsilon' & \sin\epsilon' \\
        -\sin\epsilon' & \cos\epsilon'
    \end{pmatrix}\,.
\end{equation}
As we will explicitly see later, the angle $\epsilon'$ is unphysical because it corresponds to the freedom of choosing the basis of $(B_\mu, B'_\mu)$.
Furthermore, we can set $\eta(\mu_0)=0$ at a given energy scale $\mu_0$ by redefining $y$ and $z$ as well as $g_y$ and $g_z$ at that scale.
However, as the charges should be scale independent, this redefinition can only be done at $\mu=\mu_0$ and $\eta(\mu) \ne 0$ at other scales $\mu \ne \mu_0$.

In the region of the parameter space where $g_z\ll g_y$, the coupling matrix $\mathbf g$ has the typical sizes of
\begin{equation}
\mathbf{g} = g_y \begin{pmatrix}
\rO(1) & \rO(g_z/g_y) \\
\rO(g_z^2/g_y^2) & \rO(g_z/g_y)\,.
\end{pmatrix}
\end{equation}
and thus $\eta=\rO(1)$.
In our particle spectrum shown in \tab{tab:ChargeAssignment}, if we fix $\eta=0$ at $\mu_0=10^{18}\GeV$, its low-energy value becomes $\eta(100\GeV)\simeq0.656$.
Our predictions are not much affected by the value of $\eta$, and therefore we presented the results for $\eta=0$ in the main text.
The effect of $\eta$ on our predictions are further discussed in \app{sec:appendixB}.

It is straightforward to check that $\eta$ is equivalent to the gauge kinetic mixing $\hat\epsilon$, which appears in
\beq
\bsp
\mathcal L&\supset
-\frac14 \hat F^{\mu\nu}\hat F_{\mu\nu}
-\frac14 \hat F'^{\mu\nu}\hat F'_{\mu\nu}
-\frac{\hat\epsilon}{2}\hat F^{\mu\nu}\hat F'_{\mu\nu}
,\\
 \mathcal{D}^{\mathrm{U}(1)}_\mu &= -\ri
(y \hat g_y \hat B_\mu + z \hat g_z \hat B'_\mu).
\esp
\eeq
The parameters of the two approaches are related by
 $g_y = \hat g_y$,
 $g_z = \hat g_z/\sqrt{1-\hat\epsilon^2}$, and
 $\eta = \hat\epsilon \hat g_y/\hat g_z$.

A convenient way of taking into account the effect of kinetic mixing is to define a scale-dependent effective charge~\cite{Holdom:1985ag}
\begin{equation}
 \zeta(\mu) = z-\eta(\mu) y.
\end{equation}
The $\gSU(2)\w L\otimes \UOne_y\otimes\UOne_z$ gauge sector before the spontaneous symmetry breaking is summarized by, with this effective charge,
\begin{equation}
 \mathcal{D}_\mu = -\ri \begin{pmatrix}
    y g_y & \zeta g_z
    \end{pmatrix}
    \begin{pmatrix}
        \cos\epsilon' & \sin\epsilon' \\
        -\sin\epsilon' & \cos\epsilon'
    \end{pmatrix}
    \begin{pmatrix}
    B_\mu \\ B'_\mu
    \end{pmatrix}
-\ri g\w L {\sum_i} T^i W^i_\mu\,,
\end{equation}
where $g\w L$, $T^i$, and $W^i_\mu$ respectively denotes the coupling constant, charge, and gauge boson of $\gSU(2)\w L$.

The $\gSU(2)\w L\otimes \UOne_y\otimes\UOne_z$ gauge breaks into the electromagnetic $\UOne\w{em}$ by the vacuum expectation values
\begin{equation}
 \vev{\phi_0}=\frac{v}{\sqrt2},\qquad
 \vev{\chi}=\frac{w}{\sqrt2}.
\end{equation}
The neutral gauge bosons mix into two massive and one massless gauge bosons, which we describe by
\begin{equation}
    \begin{pmatrix}
        B_\mu \\ W^3_\mu \\ B'_\mu 
    \end{pmatrix} = \mathbf{R}
    \begin{pmatrix}
        A_\mu \\ Z_\mu \\ Z'_\mu
    \end{pmatrix},
\end{equation}
with the mixing matrix containing three angles
\begin{align}
\mathbf{R}&= \mathbf{R}_{\epsilon''}\mathbf{R}_{\tW}\mathbf{R}_{\theta_Z}^\dagger
\notag\\&=
    \begin{pmatrix}
        \cos\epsilon'' & 0 & \sin\epsilon'' \\
        0  & 1 & 0 \\
        -\sin\epsilon'' & 0 & \cos\epsilon''
    \end{pmatrix}
    \begin{pmatrix}
        \cos\tW & -\sin\tW & 0 \\
        \sin\tW & \cos\tW & 0 \\
        0 & 0 & 1
    \end{pmatrix}
    \begin{pmatrix}
        1 & 0 & 0 \\
        0 & \cos\tZ & \sin\tZ \\
        0 & -\sin\tZ & \cos\tZ
    \end{pmatrix}\,.
\end{align}
The angles are fixed by requiring that the mass matrix of the gauge bosons becomes diagonal with $A_\mu$ being the massless photon.

The gauge boson mass terms emerge from $\left|\mathcal{D}_\mu\phi\right|^2 + \left|\mathcal{D}_\mu\phi\right|^2$.
Explicitly, it is given by
\begin{equation}
 \mathcal L\supset
\pmat{B^\mu\\W^{3\mu}\\B'^\mu}^\TT
\mathbf{R}_{\epsilon'}^\TT\left[
\frac{v^2}{2}
    \begin{pmatrix}
         {g_y^2}/{4}     & -{g_\rL g_y}/{4} &  {\zeta_\phi g_y g_z}/{2} \\
        -{g_\rL g_y}/{4} &  {g_\rL^2}  /{4} & -{\zeta_\phi g_\rL g_z}/{2} \\
         {\zeta_\phi g_y g_z}/{2} & -{\zeta_\phi g_\rL g_z}/{2} & (\zeta_\phi^2 + {w^2}/{v^2})g_z^2
    \end{pmatrix}
\right]
\mathbf{R}_{\epsilon'}
\pmat{B_\mu\\W^3_\mu \\B'_\mu}
\end{equation}
with
\begin{equation}
  \mathbf{R}_{\epsilon'}= \begin{pmatrix}
        \cos\epsilon' & 0 & \sin\epsilon' \\
        0  & 1 & 0 \\
        -\sin\epsilon' & 0 & \cos\epsilon'
    \end{pmatrix}\,.
\end{equation}
Therefore, one can obtain the mixing angles and the mass by solving
\begin{equation}
 \frac12\diag(0, M_Z^2, M_{Z'}^2)=
 \mathbf{R}^\TT \mathbf{R}_{\epsilon'}^\TT
 \left[ \frac{v^2}{2}
    \begin{pmatrix}
         {g_y^2}/{4}     & -{g_\rL g_y}/{4} &  {\zeta_\phi g_y g_z}/{2} \\
        -{g_\rL g_y}/{4} &  {g_\rL^2}  /{4} & -{\zeta_\phi g_\rL g_z}/{2} \\
         {\zeta_\phi g_y g_z}/{2} & -{\zeta_\phi g_\rL g_z}/{2} & (\zeta_\phi^2 + {w^2}/{v^2})g_z^2
    \end{pmatrix}
 \right]
 \mathbf{R}_{\epsilon'}\mathbf{R}\,.
\end{equation}
The mixing angles are given by
\begin{equation}
    \epsilon''=-\epsilon',\quad
    \tan\tW = \frac{g_y}{g_\rL},\quad \tan2\tZ = \frac{\zeta_\phi (g_z/g_{Z^0})}{1/4 - (\zeta_\phi^2 + w^2/v^2)(g_z/g_{Z^0})^2}\,,
\end{equation}
where $g_{Z^0}=\sqrt{g_\rL^2 + g_y^2} = g_\rL/\cW$. {We see that} the unphysical parameter $\epsilon'$ is canceled with the rotation $\epsilon''$.
The neutral gauge boson masses are given by
\begin{equation}
    M_{Z}=\frac{g_{Z^0} v}2  \left(1+\frac{2\zeta_\phi g_z}{g_{Z^0}}\tan\tZ\right)^{1/2},\quad
    M_{Z'}=g_z w
    \left(1+\frac{2\zeta_\phi g_z}{g_{Z^0}}\tan\tZ\right)^{-1/2}    \label{eq:MZ-MZ'}\,,
\end{equation}
{while} the $W$-boson mass is given by $M_W = g\w L v/2$.

{Finally, the effective couplings in Eq.~\eqref{eq:effcoups} can straightforwardly be derived from the neutral part of the covariant derivative}
\begin{align}
 \mathcal{D}_\mu^{\text{neutral}}
&= -\ri \begin{pmatrix}
    y g_y & T_3 g\w L & \zeta g_z
    \end{pmatrix}
    \mathbf{R}_{\tW}\mathbf{R}_{\theta_Z}^\dagger
    \begin{pmatrix}
        A_\mu \\ Z_\mu \\ Z'_\mu
    \end{pmatrix}.
\end{align}

\section{Dependence on the parameter \texorpdfstring{$\eta$}{eta}}
\label{sec:appendixB}

For fermions with non-vanishing electric charge, the vertex $Z'$--$f$--$\bar f$ depends on the gauge-mixing parameter $\eta$ (see Eq.~\eqref{eq:Dmu-neutral} with Eq.~\eqref{eq:QZp-vec}). For charged leptons, in particular for the electron, the modified vertex reads
\begin{equation}
    \Gamma^\mu_{Z'ee}(\eta)
    =-\ri\gamma^\mu \mathcal {Q}_{Z'}^{e}
    \simeq-\ri g_z\gamma^\mu\left[(\eta-2)\cos^2\tW+\frac{1}{2}\right]\,.
\end{equation}
The introduction of the extra parameter affects both the freeze-in and freeze-out dark matter production mechanisms. However, the change in the allowed parameter space is relatively small, and the results are qualitatively the same.

\begin{figure}
    \centering
    \includegraphics[width=0.7\linewidth]{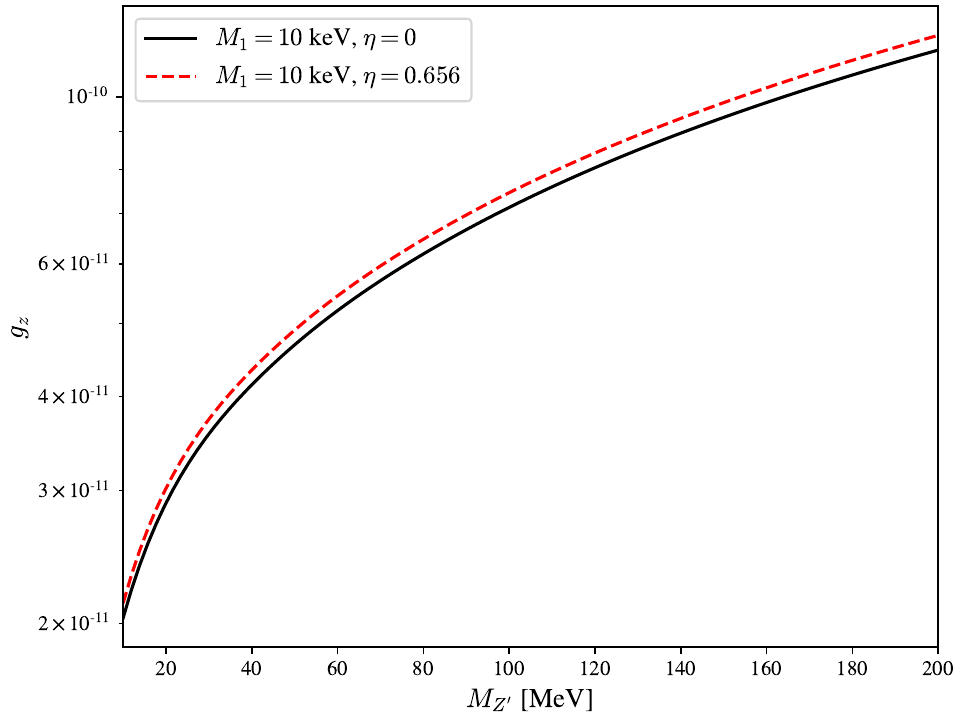}
    \caption{Comparison of the super-weak couplings $g_z$ reproducing dark matter densities for $N_1$ for the extremal values of the gauge mixing $\eta$. The difference in the couplings is described exactly by Eq.~\eqref{eq:freezeinetacompare}.}
    \label{fig:FreezeInCompare}
\end{figure}

In the freeze-in case, the two important quantities we have to look at are the branching ratio of $Z'$ bosons into right-handed neutrinos $\mathcal{B}^{Z'}_{N_1N_1}$, and the decay rate of $Z'$ bosons into standard model particles ($\Gamma_\mathrm{SM}$). By virtue of the production mechanism, the relic density of $N_1$ is
\begin{equation}
    \mathcal{Y}_\infty = \mathcal{B}^{Z'}_{N_1N_1}(\eta;M_1,M_{Z'})\cdot\mathrm{max}\big(\mathcal{Y}_{Z'}\big)\,,
\end{equation}
where the maximum of the $Z'$ comoving number density will be proportional to $\Gamma_\mathrm{SM}$ because the $Z'$ bosons are produced via the inverse decays of electrons and standard model neutrinos. In conclusion, for a fixed value of $M_1$ and $M_{Z'}$, we can connect the required couplings for $\Omega_\text{DM}$ with $\eta$ as
\begin{equation}
    \label{eq:freezeinetacompare}
    g_z(\eta_2) = g_z(\eta_1)\sqrt{\frac{\mathcal{B}^{Z'}_{N_1N_1}(\eta_1)\tilde{\gamma}_\mathrm{SM}(\eta_1)}{\mathcal{B}^{Z'}_{N_1N_1}(\eta_2)\tilde{\gamma}_\mathrm{SM}(\eta_2)}}\,,
\end{equation}
where $\tilde{\gamma}_\mathrm{SM}=\Gamma_\mathrm{SM}/(M_{Z'}g_z^2)$.
An example comparison is shown in Fig~\ref{fig:FreezeInCompare}.

Contrary to the freeze-in case, the $\eta$-dependence of freeze-out is less trivial, but regardless it can be largely neglected. 
To showcase the difference explicitly, we take a look at the solution obtained at $M_1=30\,$MeV and $\eta=0.0$ (cf. Fig. \ref{fig:ResonantFreezeOutParamSpace}), and compare it to that obtained with $\eta_\mathrm{max}=0.656$.

\begin{figure}
    \centering
    \begin{subfigure}[b]{0.48\textwidth}
        \includegraphics[width=\textwidth]{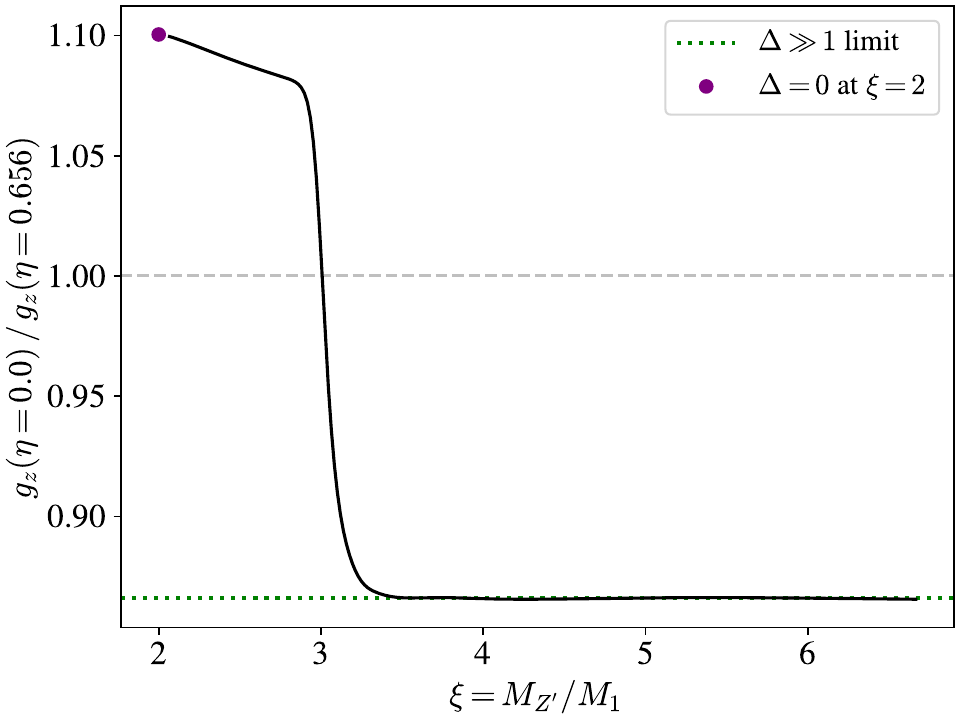}
        \caption{}
        \label{fig:CouplingsEtaCompare}
    \end{subfigure}
    ~
    \begin{subfigure}[b]{0.48\textwidth}
        \includegraphics[width=\textwidth]{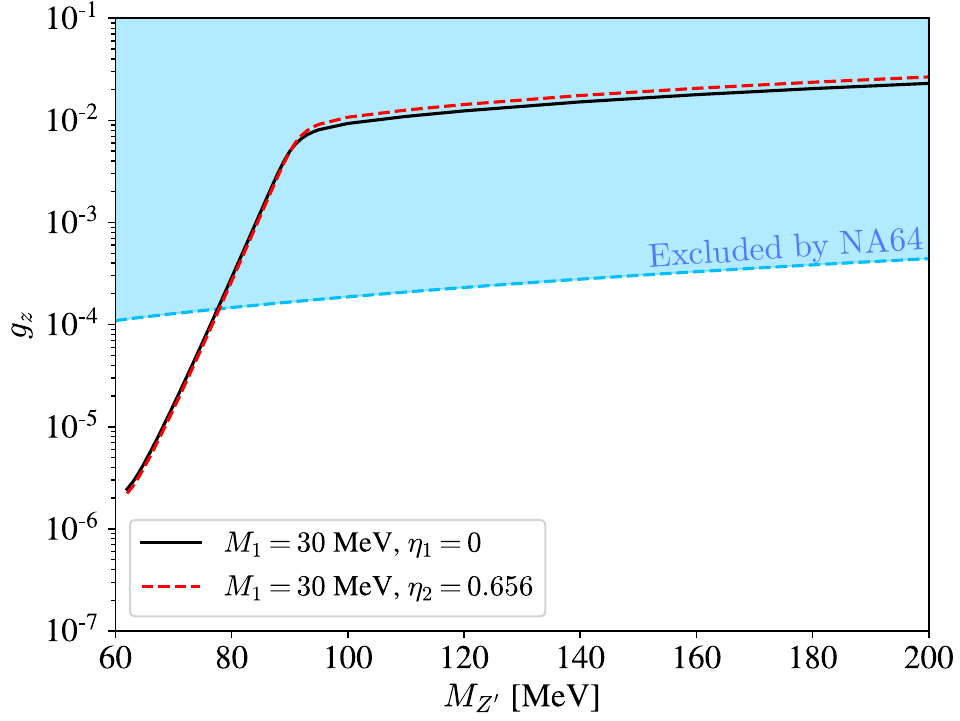}
        \caption{}
        \label{fig:ParamsSpaceEtaCompare}
    \end{subfigure}
    \caption{Comparison of the required couplings for $\Omega_\text{DM}$ in the freeze-out scenario at the extremal values of $\eta$, at $\eta_1=0$ and $\eta_2=0.656$. The value of $M_1=30\,$MeV was fixed in this example. (a) The ratio of required couplings depends on $\xi$, which is related to resonance effects. The two limits which can be determined analytically are shown: the dotted green line shows the limit of negligible resonance ($\Delta\gg 1$), while for total resonance dominance ($\Delta=0$), the purple dot indicates the endpoint. (b) Parameter space on the \{$M_{Z'}$, $g_z$\} plane for the two values of $\eta_{1,2}$.}
    \label{fig:etaCompare}
\end{figure}

In Eq. \eqref{eq:sigma_parts} we have separated the thermally averaged cross section into two parts, the resonant and the low temperature contribution. When compared to Fig. \ref{fig:ResonantFreezeOutParamSpace}, we concluded that at large values of $\xi=M_{Z'}/M_1$ the resonance is negligible, but later on it gets exponentially dominant, shown by the drastic decrease in the super-weak coupling. While $\sigmav$ depends on the masses in a complicated way, the $\eta$ and $g_z$ dependence is relatively straightforward. Schematically we write
\begin{equation}
    \sigmav(g_z,\,\eta;\,M_{Z'},\,M_1) = g_z^4 Q(\eta) \left[\frac{R(M_{Z'},\,M_1)}{\gamma(g_z,\,\eta;\,M_{Z'},\,M_1)} + L(M_{Z'},\,M_1)\right]\,,
\end{equation}
where the functions $R$ and $L$ correspond to the resonant, and low temperature contributions which may be defined via comparing to Eqs. \eqref{eq:sigma_resonant} and \eqref{eq:sigmalow}. We require that the thermally averaged cross section remains unchanged $\sigmav(g_z,\,\eta)=\sigmav(g'_Z,\,\eta')$, so that the right-hand side of the Boltzmann equation is numerically the same, then we find a connection between the ratio of the coupling $g_z$ and $\eta$. It is instructive to define $\Delta = L/R$, the relative strength of the low temperature versus the resonant contributions. Then
\begin{equation}
    \frac{g_z(\eta_1)}{g_z(\eta_2)} = \left[\frac{Q(\eta_2)\gamma(\eta_1)}{Q(\eta_1)\gamma(\eta_2)}\right]^{{1}/{4}} \left[ \frac{1+\Delta\cdot\gamma(\eta_2)}{1+\Delta\cdot\gamma(\eta_1)}\right]^{{1}/{4}} = 
    \begin{cases}
        \displaystyle\left[\frac{Q(\eta_2)}{Q(\eta_1)}\right]^{{1}/{4}}\,, & \quad \text{if }\Delta\gg 1 \\
        \displaystyle\left[\frac{Q(\eta_2)\Tilde{\gamma}(\eta_1)}{Q(\eta_1)\Tilde{\gamma}(\eta_2)}\right]^{{1}/{2}}\,, & \quad \text{if }\Delta\ll 1\,,
    \end{cases}
\end{equation}
where we now defined $\Tilde{\gamma}=\gamma/g_z^2$. While $\Delta\gg 1$ can be valid in a wide range of the parameters, i.e.,~the resonance can be fully ignored for $\xi\gtrsim 4$, the other case $\Delta\ll 1$ is only valid when $\xi\to 2$, i.e.,~when $L\to0$ due to the vanishing integral domain.

In Fig.~\ref{fig:etaCompare}, we solve the Boltzmann equation in the freeze-out case with $\eta=0$ and $\eta=0.656$, and compare the required couplings for $\Omega_\mathrm{DM}$.
Let $\eta_1<\eta_2$, with both values in the range $[0,\,0.656]$. 
By substitution into the definition of $Q(\eta)$, Eq.~\eqref{eq:Qdef}, we find $Q(\eta_1)>Q(\eta_2)$.
Similarly, for the decay rates $\Tilde{\gamma}(\eta_1)>\Tilde{\gamma}(\eta_2)$. 
It follows that for $\Delta\gg 1$ the coupling ratio ${g_z(\eta_1)}/{g_z(\eta_2)}< 1$, and by comparing $\Tilde{\gamma}(\eta)$ to $Q(\eta)$ we find that for $\Delta\ll 1$ the coupling ratio is greater than unity.
The resulting curve in Fig.~\ref{fig:CouplingsEtaCompare} follows these conclusions in the limits $\xi\to 2$ and $\xi\gtrsim 4$, while for intermediate values of $\xi$, the correspondence is not trivial, can be described with the curve shown.

We conclude, that by changing the mixing parameter $\eta$ we only slightly affect the value of the coupling $g_z$, at most by roughly 15\%.

\section{Constraints from the NA64 experiment}
\label{appendix:C}

The NA64 experiment is described in detail in Ref.~\cite{NA64:2019imj}. 
The experiment consists of an electron beam fired at a fix target of material with atomic number Z. 
The electrons interact with the target, and may emit a dark photon $A'$ via the bremsstrahlung process
\begin{equation}
    e+\mathrm{Z}\longrightarrow e+\mathrm{Z}+A'\,,\quad\text{with}\quad A'\longrightarrow\text{invisible final states.}
\end{equation}
In principle the dark photon could have visible (charged lepton pair) and invisible (neutrinos or dark particles) decay channels as well, but for simplicity in NA64 the invisible branching ratio was assumed to be unity. 
By looking for missing energy in single electromagnetic shower events, Ref.~\cite{NA64:2019imj} constrains the kinetic mixing angle $\epsilon$ of the dark photon model versus the mass of the dark photon $M_{A'}$.

The kinetic mixing angle $\epsilon$ is not a parameter of our model, indeed we use $\eta=0$ in the main text which may be translated to having $\epsilon=0$. 
However the constraints obtained for the dark photon model can be translated to constraints for the super-weak coupling $g_z$ as shown in Ref.~\cite{Ilten:2018crw}. 
We require the equality
\begin{equation}
    \label{eq:crosssectionequality}
    \sigma(e+\mathrm{Z}\longrightarrow e+\mathrm{Z}+A')~\mathcal{B}^{A'}_\mathrm{inv} =\sigma(e+\mathrm{Z}\longrightarrow e+\mathrm{Z}+Z')~\mathcal{B}^{Z'}_\mathrm{inv}
\end{equation}
where the left-hand side involves the cross section and invisible branching ratio ($\mathcal{B}_\mathrm{inv}$) calculated in the dark photon model, while the right-hand side is the same in the super-weak model. 
In the approximations outlined in the main text, the coupling of $Z'$ with fermions is vector-like and thus the comparison of the cross sections is trivial,
\begin{equation}
    \frac{\sigma(e+\mathrm{Z}\longrightarrow e+\mathrm{Z}+A')}{\sigma(e+\mathrm{Z}\longrightarrow e+\mathrm{Z}+Z')}=\frac{(e\epsilon)^2}{\Tilde{g}_z^2}
\end{equation}
where the shorthand notation $\Tilde{g}_z$ stands for
\begin{equation}
    \Tilde{g}_z = g_z\Big[(\eta-2)\cos^2\tW+\frac{1}{2}\Big]\,.
\end{equation}
The invisible branching ratio is easily obtained from Eqs.~\eqref{eq:DecayRate:Z'ee}--\eqref{eq:DecayRate:Z'N1N1}, noting that both sterile and standard model neutrinos are effectively invisible for the experimental setup. 
By setting $\eta=0$, we find
\begin{equation}
    \mathcal{B}^{Z'}_\mathrm{inv}=\frac{\displaystyle \sum_{i=1}^{3}\Gamma(Z'\to\nu_i\nu_i) + \Gamma(Z'\to N_1N_1)}{\displaystyle \sum_{i=1}^{3}\Gamma(Z'\to\nu_i\nu_i) + \Gamma(Z'\to N_1N_1)+\Gamma(Z'\to ee)} = \frac{\displaystyle 3+\left(1-\frac{4M_1^2}{M_{Z'}^2}\right)^{3/2}}{\displaystyle 11.887 + \left(1-\frac{4M_1^2}{M_{Z'}^2}\right)^{3/2}}\,,
\end{equation}
which depends on the mass ratio weakly through $\Gamma(Z'\to N_1N_1)$. 
For simplicity we use the most conservative estimate for the invisible branching ratio, i.e.,~when its value is the largest, at $M_{1}\ll M_{Z'}$. 
The reason for doing so is to eliminate the mass dependence, while providing the most stringent cut on the parameter space available. 
Thus we use
\begin{equation}
    \mathcal{B}^{Z'}_\mathrm{inv} = 0.3104 - \rO\left(\frac{M_1^2}{M_{Z'}^2}\right).
\end{equation}
Substitution into Eq.~\eqref{eq:crosssectionequality} with $\eta=0$ and ignoring the mass correction to the invisible branching ratio leads to
\begin{equation}
    \epsilon=\frac{|\Tilde{g}_z|}{e}\sqrt{\mathcal{B}^{Z'}_\mathrm{inv}}\simeq 1.94~g_z\,.
\end{equation}
This means that by ignoring the weak mass-dependence, a simple linear rescaling of the coupling is needed for including the NA64 exclusion on the $M_{Z'}$--$g_z$ plane. 
We note that fixing the mass ratio to the resonance condition $M_{Z'}=2M_1$ would have produced a 10\% difference in the scaling, which can be safely ignored at our precision for the sake of simplicity.

\providecommand{\href}[2]{#2}\begingroup\raggedright\endgroup

\end{document}